\documentclass{aastex631}

\usepackage{comment}
\usepackage{rotating}
\usepackage{booktabs}

\received{July 24, 2023}
\revised{November 15, 2023}
\accepted{November 18, 2023}

\submitjournal{ApJ}

\shorttitle{Alfvénic Fluctuations and Magnetic Coherence in ICMEs}
\shortauthors{Scolini et al.}

\graphicspath{{./}{Figures/}}

\begin{document}

\title{On the Role of Alfvénic Fluctuations as Mediators of Coherence within Interplanetary Coronal Mass Ejections:
Investigation of Multi-Spacecraft Measurements at 1~au}

\correspondingauthor{Camilla Scolini}
\email{camilla.scolini@gmail.com}

\author[0000-0002-5681-0526]{Camilla Scolini}
\affiliation{Institute for the Study of Earth, Oceans, and Space, University of New Hampshire, Durham, NH, USA}

\author[0000-0002-1890-6156]{No\'e Lugaz}
\affiliation{Institute for the Study of Earth, Oceans, and Space, University of New Hampshire, Durham, NH, USA}

\author[0000-0002-9276-9487]{R\'eka M. Winslow}
\affiliation{Institute for the Study of Earth, Oceans, and Space, University of New Hampshire, Durham, NH, USA}

\author[0000-0001-8780-0673]{Charles J. Farrugia}
\affiliation{Institute for the Study of Earth, Oceans, and Space, University of New Hampshire, Durham, NH, USA}

\author[0000-0001-5731-8173]{Norbert Magyar}
\affiliation{Centre for mathematical Plasma Astrophysics, Department of Mathematics, KU Leuven, Leuven, Belgium}

\author[0000-0002-7526-8154]{Fabio Bacchini}
\affiliation{Centre for mathematical Plasma Astrophysics, Department of Mathematics, KU Leuven, Leuven, Belgium}
\affiliation{Royal Belgian Institute for Space Aeronomy, Solar-Terrestrial Centre of Excellence, Uccle, Belgium}

\begin{abstract}
Interplanetary coronal mass ejections (ICMEs) are defined as ``coherent'' if they are capable of responding to external perturbations in a collective manner. This implies that information must be able to propagate across ICME structures, and if this is not the case, single-point in-situ measurements cannot be considered as indicative of global ICME properties. 
Here, we investigate the role of Alfvénic fluctuations (AFs) as mediators of ICME coherence. We consider multi-point magnetic field and plasma measurements of 10 ICMEs observed by the ACE and Wind spacecraft at 1~au at longitudinal separations of $0.5^\circ-0.7^\circ$. For each event, we analyze the Alfvénicity in terms of the residual energy and cross helicity of fluctuations, and the coherence in terms of the magnetic correlation between Wind and ACE.
We find that $\sim65$\% and 90\% of ICME sheaths and magnetic ejecta (MEs), respectively, present extended AFs covering at least 20\% of the structure. 
Cross helicity suggests AFs of solar and interplanetary origin may co-exist in the ICME population at 1 au. 
AFs are mainly concentrated downstream of shocks and in the back of MEs. 
The magnetic field is poorly correlated within sheaths, 
while the correlation decreases from the front to the back of the MEs for most magnetic field components. 
AFs are also associated with lower magnetic field correlations. This suggests either that ICME coherence is not mediated by Alfv\'en waves, implying that the coherence scale may be smaller than previously predicted, or that the magnetic field correlation is not a measure of coherence.
\end{abstract}

\keywords{Solar coronal mass ejections (310) --- Solar wind (1534) --- Interplanetary magnetic fields (824) --
Alfv\'en waves (23)}


\section{Introduction} 
\label{sec:introduction}

Interplanetary coronal mass ejections \citep[ICMEs;][]{Bothmer1998, Cane2003, Kilpua2017} are said to be ``coherent'' if they are capable of responding to external perturbations in a collective manner \citep{Burlaga1981, Owens2017}. Whether ICMEs behave coherently at global or only at local scales is a topic of debate \citep[see e.g.][]{Lugaz2018, Owens2020, Al-Haddad2022, Scolini2023} due to its implications for the global evolution of ICME magnetic structures and the interpretation of single-point in-situ measurements. 
Throughout this work, we use the term ``magnetic ejecta'' \citep[ME;][]{Winslow2015} to refer to the magnetically dominated portion of an ICME, which is identified by an enhanced magnetic field and low levels of magnetic fluctuations compared to the preceding and following interplanetary magnetic field. This definition of ME includes as subset the structures called ``magnetic clouds'' \citep[MCs;][]{Burlaga1981}, which require the presence of smoothly rotating magnetic fields, low density and low temperature in addition to the ME signatures. Further clarifications on how this choice of terminology applies to the specific events considered in this study are provided below.


To date, a number of complementary approaches have been pursued to shed light on the scale of coherence within ICMEs, yielding partial, and sometimes inconsistent, results.\
On the one hand, early observations by \citet{Burlaga1981}, for example, presented evidence that ICMEs can exhibit comparable plasma and magnetic field properties across a longitudinal separation of $\sim 30^\circ$ at least. 
Other multi-spacecraft measurements indicate that some of the ICME properties may be global, and that a comparable magnetic configuration can be measured within magnetic ejecta (MEs) by spacecraft at separations $\ge 30^\circ$ \citep[e.g.,][]{Cane1997, Kilpua2011, Good2016, Lugaz2022}. Evidence in favor of a global coherence of ICME structures is well represented by the famous cartoon by \citet{Zurbuchen2006} (see Figure~2 therein), of which several variations have since been proposed \citep[as summarized in Figure~1 by][]{Owens2016}.
On the other hand, based on considerations on the shape, expansion speeds, and Alfvén speed of ICMEs, \citet{Owens2017} estimated that ICMEs (and particularly their MEs) may cease to behave as globally coherent structures at heliocentric distances of $0.2-0.5$~au, and maintain a coherent scale of around $26^\circ$ by the time they reach 1 au. Such an estimate is intimately linked to the assumption that ICMEs maintain their angular width during propagation, in such a way that non-radial flows should not be present, neither away nor towards the ICME axis. 
\citet{Al-Haddad2022} challenged such a scenario revealing the presence, at least in two case studies observed at 1 au, of small non-radial expansion rates consistent with an ICME cross section more elliptical than implied by the kinematic model assumed by \citet{Owens2017}. If applicable to the general ICME population, such results may indicate that at 1 au, ICME coherence could only be sustained at even smaller scales than previously estimated. However, no steady, large non-radial flows inside MEs were detected over a set of 48 ICMEs considered by \citet{Al-Haddad2022}, which is more consistent either with an ICME cross section that becomes highly elliptical due to kinematic effects, or with a scenario where the entire paradigm of MEs described in terms of flux ropes (FR) with an axial invariance might be too limited. Importantly, \citet{Al-Haddad2022} also raised the issue of distinguishing between ICME lateral expansion and deflection from single-point measurements, stressing the need for multi-point observations at two opposite sides of an ICME to advance our understanding of ICME coherence from an observational standpoint. 

In addition to the scarcity of multi-point ICME observations, estimating the scale of ICME coherence is further complicated by the fact that coherence is not a physical observable, and as such, assumptions have to be made when interpreting observational data with respect to the coherence of ICME structures. Observationally, the correlation of magnetic field components as measured by different spacecraft crossing the same ICME through different trajectories has been commonly assumed as a proxy for ICME coherent behavior \citep{Matsui2002, Farrugia2005, Lugaz2018, Ala-Lahti2020}. 
The investigation by \citet{Lugaz2018} suggested the existence of two characteristic scales of correlation within ICMEs near 1 au: one related to the magnetic field components (around $4^\circ-7^\circ$) and one for the total magnetic field (around $14^\circ-20^\circ$). However, so far limited multi-point ICME observations prevented a direct and systematic exploration of the parameter space, particularly at scales between $1^\circ$ and $10^\circ$. Additionally, such observational studies have been so far only possible near 1 au due to the lack of spacecraft reaching close angular separations at inner heliocentric distances prior to the current solar cycle. In an attempt to overcome these observational limitations, in \citet{Scolini2023} we performed 3-D numerical simulations of ICMEs in the inner heliosphere which highlighted the role of interactions with other large-scale structures, such as high speed streams (HSSs) and stream interaction regions (SIRs), as a primary mechanism acting to decrease the correlation scale of the magnetic field components within ICMEs. This study revealed how, in such cases, the correlation is progressively lost by ICMEs during propagation between 0.1 and 2 au. 

Overall, while previous works agree in considering ICMEs, and particularly MEs, as coherent structures only at scales around or smaller than $\sim 25^\circ$ at 1 au, how this depends on the ICME heliocentric distance and interaction history, remains unassessed from an observational standpoint. At the same time, the fundamental nature of the relationship between ICME coherence and the correlation of magnetic field time profiles measured across different locations within the same ICME remains unexplored, and to be answered, requires first of all a physical understanding of how information about the acting external forces is transmitted across ICME structures.

\smallskip
One fundamental question with respect to the coherence of ICMEs relates to the nature of the information carrier required to propagate information about an external perturbation across an ICME structure. In fact, the very definition of magnetic coherence implies that a coherent behavior can only be exhibited if information about the acting perturbation is able to propagate across an ICME structure, but the nature of such an information carrier is still an open question in its own right. 
In this paper, we explore the possibility that Alfv\'en waves \citep{Alfven1942} are the main mediators of coherent behavior across ICME structures. 
In the theory of magnetohydrodynamics (MHD), Alfv\'en waves consist of low-frequency \citep[i.e. much lower than the ion cyclotron frequency, which is typically around $10^{-2}-10^{0}$~Hz 
in the solar wind at 1 au and inner heliocentric distances; e.g.][]{Bale2016, Zhao2021}, non-compressive (shear), in-phase oscillations of plasma and magnetic field lines perpendicular to the local magnetic field direction, propagating along magnetic field lines. 
In the solar wind, Alfv\'en waves are generally called Alfv\'enic fluctuations (AFs).
AFs are extensively present in both the solar atmosphere and the solar wind \citep[see, for example,][]{Belcher1969, DePontieu2007, Tomczyk2007, Wang2012, DAmicis2021} and can be generated through a variety of mechanisms including magnetic reconnection \citep[e.g.][]{Kigure2010, Cranmer2018} and various changes in the force-balanced state of a magnetic flux-rope structure \citep[e.g.][]{Hollweg1982, Velli1999, Longcope2000}. Particularly in the solar wind, AFs are observed to propagate predominantly along the anti-sunward direction, suggesting AFs of solar origin are weakly damped in the interplanetary medium and can efficiently propagate up to 1~au and beyond \citep{Belcher1971, Chen2016}. 
Thus, our intuition is that due to their ubiquitous presence in the solar wind and the broad range of physical mechanisms able to generate them, AFs are the most prominent candidates for the propagation of information throughout ICME structures. Such a hypothesis was also implicitly suggested by \citet{Owens2017, Owens2020}, but was not followed up by any quantitative verification based on theoretical nor observational evidence.
General discussions on the role of AFs as propagators of information in other contexts in space plasma can be also found in classic textbooks, see, e.g., \cite{Kivelson1995}.
Only in recent years have large-amplitude (comparable to the average background magnetic field) 
AFs within MEs been first detected in Helios data between 0.3~au and 0.7~au \citep{Marsch2009, Yao2010}. A handful of studies \citep[e.g.][]{Yao2010, Liang2012} related AFs within MEs to solar formation mechanisms, particularly in relation to solar prominence eruptions. Solar prominences are common progenitors of ICMEs \citep[e.g.,][]{StCyr1991, Scolini2018} and routinely exhibit the presence of MHD waves and oscillations \citep[e.g.,][]{Okamoto2007, Arregui2018}, which suggests many CMEs may be filled with AFs already in the solar corona. On the other hand, several studies also highlighted a general scarcity of AFs within prominence-erupted ICMEs compared to the solar corona, suggesting AFs tend to quickly dissipate as ICMEs propagate away from the Sun \citep{Marsch2009, Yao2010, Li2016, Li2017}. Another possibility, yet less explored, is given by interplanetary formation of AFs within MEs via ubiquitous plasma processes such as magnetic reconnection \citep[e.g.][]{Gosling2005} and/or the interaction of plasma structures with velocity shears \citep[e.g.][]{Bavassano1978, Roberts1992}.

\smallskip
Based on the above discussion, the goals of this study are twofold: 
first, we want to characterize and quantify the Alfv\'enicity of ICMEs at 1 au. 
Second, we want to answer the question of whether AFs are significant mediators of coherent behavior within ICME structures, in a scenario where a coherent behavior is measured by the correlation of magnetic field profiles across different ICME locations.
In other words, we want to understand if and how AFs alter the internal structure of ICMEs at large to ``meso'' (intermediate between the ICME size and kinetic) scales.
In this first study on the topic, we aim to answer these questions by uncovering general trends that exist within the ICME population. Such an approach requires the consideration of a homogeneous set of ICMEs observed by multiple spacecraft at comparable heliocentric distances and angular separations, and for which multi-point in-situ plasma and magnetic field data are available. 
At the time of writing, such a data set is only available at 1 au. 
Therefore, in this study we specifically consider a set of 10 ICMEs observed near Earth by the ACE and Wind spacecraft at longitudinal separations between 0.5$^\circ$ and $0.7^\circ$. 
We investigate the Alfv\'enicity of their plasma and magnetic field fluctuations in two regimes of the fluctuations' power spectrum: (i) the injection range, covering wavenumbers of the largest-scale fluctuations, excited by macroscopic dynamics, such as prominence oscillations in the corona \citep[typically covering frequencies $<10^{-3}$~Hz for MEs at 1 au;][]{Good2022}, and (ii) the inertial range, which corresponds to the range of wavenumbers where self-similar (fluid-like, i.e. MHD) cascades transport the energy injected at the injection scales towards progressively smaller scales \citep[typically covering frequencies at $10^{-3} - 10^{-2}$~Hz for MEs at 1 au;][]{Good2022}.
In this study, we focus specifically on frequencies between $2.3 \times 10^{-5} - 3.3 \times 10^{-3}$~Hz (corresponding to temporal scales between 12 hours and 5 minutes), as they cover the injection range and low-frequency end of the inertial range, which correspond to the large-to-meso ICME scales.
Additionally, we investigate the correlation of magnetic field profiles at 0.5$^\circ$ to $0.7^\circ$ of angular separation in order to draw general conclusions regarding the role of AFs as mediators of ICME coherence and magnetic field correlation. 

The paper is structured as follows.
Section~\ref{sec:data_and_methods} provides an overview of the data sets and methods used to quantify the Alfv\'encity and correlation between in-situ ICME properties at different spacecraft locations. 
Section~\ref{sec:results} presents the results of the analysis of a set of 10 ICMEs detected at 1 au.  
In Section~\ref{sec:conclusions} we summarize our results and discuss them in the context of understanding the relationship between coherence and AFs within ICMEs near 1~au. 

\section{Data and methods}
\label{sec:data_and_methods}

\subsection{Data}
\label{subsec:data}

We start our event selection from the list of 35 ICMEs observed by ACE and Wind at 1~au analyzed by \citet{Lugaz2018}.
All these ICMEs present MC or MC-like signatures \citep[i.e. listed as ``2'' or ``1'' in][respectively]{Richardson2010}. Thus, we refer to the ejecta part of these ICMEs as MEs, whether they are MCs or MC-like.
We filter out events observed when the two spacecraft had longitudinal separation larger than $0.5^\circ$ (corresponding to $\sim 200$ Earth radii at 1~au) in order to ensure the observing spacecraft sampled the ICME structures across sufficiently different directions (these separations correspond to an Alfv\'en propagation time of about 4 hours between the spacecraft). 
Previous studies established that the typical ME duration at 1 au ranges between 18 and 26 hours \citep{Gopalswamy2015, NievesChinchilla2018}, while small flux ropes (SFRs) can have duration up to 12 hours \citep{Yu2014}. To ensure sample homogeneity and prevent contamination, we thus discard all ICMEs that have an ME duration of less than 12 hours at 1 au, and that lack a clear solar counterpart.\ 
Such a choice was based on the fact that weak and slow interplanetary magnetic flux ropes such as SFRs may be ascribed to other formation mechanisms than large-scale solar eruptions \citep[e.g., through streamer blow-outs or interplanetary reconnection;][]{Moldwin1995, Cartwright2010, Sanchez-Diaz2017, Lavraud2020}.\ Due to their different formation and propagation histories, they can therefore be expected to yield a different Alfv\'enic content at 1 au compared to ICMEs, particularly if AFs are formed during solar eruptive events such as in relation to prominence eruptions.

This selection criteria results in 10 ICMEs observed at longitudinal separations between $0.5^\circ$ and $0.7^\circ$ (corresponding to $\sim 200 - 300$ Earth radii, or $\sim 0.009 - 0.013$~au, at 1~au), so that the typical Alfv\'en travel time within the ICME between the two spacecraft locations was typically around or larger than 4~hours. As a result of these selection criteria, 9 out of the 10 ICMEs selected were propagating fast enough to have driven a shock and sheath by the time they reached 1~au.
For each event, we analyze the Alfv\'{e}nicity within the ICMEs using the wavelet analysis described in Section~\ref{subsubsec:methods_af_identification} below, and the correlation between the magnetic field time series at Wind and ACE using the methodology described in Section~\ref{subsubsec:methods_correlation} below. 

For each ICME driving a shock, we cross-check our identifications of the ICME start time with the Heliospheric Shock Database, generated and maintained at the University of Helsinki \citep[][\url{http://ipshocks.fi}]{Kilpua2015}.
For each event in the list, we determine the ME boundaries as follows. 
The ME start and end times are initially selected based on the boundaries listed in the HELIO4CAST ICMECAT \citep[][\url{https://helioforecast.space/icmecat}]{Moestl2017} and Wind ICME \citep[][\url{https://wind.nasa.gov/ICME_catalog/ICME_catalog_viewer.php}]{NievesChinchilla2018} catalogs, but are then adapted visually as follows. First, Wind and ACE magnetic field and plasma measurements are investigated separately, and ME boundaries are chosen independently at the two spacecraft. In case of data gaps in the magnetic field and/or plasma data right where one of the boundaries is expected to occur (as in the case of ACE plasma data around the ME end boundary for Event 6), features at the other spacecraft are used to guide the selection of the ME boundaries at the spacecraft affected by the data gaps. As a result, the ME boundaries are chosen as consistently as possible between ACE and Wind.
These ICME boundaries are used for the rest of the analysis, and are provided in Table~\ref{tab:icmes}.

Throughout our analysis, we use ACE magnetic field data at 16-s cadence taken by the Magnetic Field Experiment \citep[MAG;][]{Smith1998}, and measurements of the solar wind plasma and suprathermal electron properties at 64-s cadence from the Solar Wind Electron Proton Alpha Monitor \citep[SWEPAM;][]{McComas1998}. 
At Wind, we use measurements of the magnetic field at 1-min cadence taken by the Magnetic Field Investigation \citep[MFI;][]{Lepping1995}, complemented by measurements of the solar wind plasma properties at 3-s cadence, and of suprathermal electron pitch angle distribution data at 24-s cadence from the Three-Dimensional Plasma and Energetic Particle Investigation \citep[3DP;][]{Lin1995}. 
When appropriate, we compare results obtained from the Wind/3DP data set against results obtained by considering solar wind plasma properties measured at 92-s cadence from the Solar Wind Experiment \citep[SWE;][]{Ogilvie1995}.
The magnetic field and velocity data used in this paper are given in RTN coordinates.

\begin{table}[]
\centering
\begin{tabular}{l|ccc|ccc}
 \hline
 \hline
                                          & & Wind & & & ACE & \\
 \hline
 Event no.                                & ICME start time & ME start time & ME end time & ICME start time & ME start time & ME end time \\
 \hline

 1 & 2000-10-12 22:33 &	2000-10-13 17:24 & 2000-10-14 19:12 &  2000-10-12 21:45 & 2000-10-13 16:10 & 2000-10-14 17:20 \\
 2 & 2001-03-27 18:07 &	2001-03-27 20:00 & 2001-03-28 14:23 &  2001-03-27 17:15 & 2001-03-27 19:00 & 2001-03-28 13:40 \\
 3 & 2001-04-11 14:09 &	2001-04-11 22:48 & 2001-04-12 17:58 &  2001-04-11 13:14 & 2001-04-11 22:20 & 2001-04-12 17:25 \\
 4 & 2001-04-21 15:29 &	2001-04-22 00:10 & 2001-04-23 01:11 &  2001-04-21 15:06 & 2001-04-21 23:30 & 2001-04-23 00:15 \\
 5 & 2001-04-28 05:00 &	2001-04-28 14:50 & 2001-04-29 16:00 &  2001-04-28 04:31 & 2001-04-28 14:00 & 2001-04-29 15:00 \\
 6 & 2001-07-10 10:30 &	2001-07-10 10:30 & 2001-07-11 11:00 &  2001-07-10 09:18 & 2001-07-10 09:18 & 2001-07-11 08:35 \\
 7 & 2001-12-29 05:17 &	2001-12-30 01:00 & 2001-12-30 19:09 &  2001-12-29 04:47 & 2001-12-30 00:10 & 2001-12-30 18:30 \\
 8 & 2002-02-28 05:06 &	2002-02-28 19:11 & 2002-03-01 23:15 &  2002-02-28 04:00 & 2002-02-28 16:55 & 2002-03-01 23:15 \\
 9 & 2002-05-18 19:45 &	2002-05-19 03:30 & 2002-05-20 03:34 &  2002-05-18 19:19 & 2002-05-19 03:00 & 2002-05-20 02:50 \\
 10 & 2002-05-20 03:35 &	2002-05-20 11:15 & 2002-05-21 21:00 &  2002-05-20 02:57 & 2002-05-20 11:00 & 2002-05-21 20:30 \\
 \hline
 \end{tabular}
 \caption{Summary of the ICME times at ACE and Wind.}
\label{tab:icmes}
\end{table}

\subsection{Methods}
\label{subsec:methods}

\subsubsection{Characterization of Alfv\'enicity} 
\label{subsubsec:methods_af_identification}

To identify periods of AFs in and around ICMEs (including MEs, preceding sheaths, and surrounding solar wind), velocity and magnetic field fluctuations are explored through continuous wavelet spectrograms of the normalized residual energy 
\begin{equation}
    \sigma_r(k,t)=\frac{E_v(k,t)–E_B(k,t)}{E_v(k,t)+E_B(k,t)},
\end{equation}
where $E_v(k,t)$ and $E_B(k,t)$ are the sum of the power of the wavelet transforms \citep{Torrence1998} of the components of the velocity $\vec{v}(t)$ and magnetic field $\vec{B}(t)$ vectors, respectively \citep[][]{Telloni2012, Telloni2013, Telloni2021, Good2020, Good2022}, and are functions of time $t$ and of the wavenumber $k$. 
$\sigma_r(k,t)$ measures the imbalance between the kinetic and magnetic energies and is expected to be close to zero in a reference frame co-moving with the solar wind due to the equipartition of magnetic and kinetic energy of AFs. 
This method enables us to investigate large periods of data through visual inspection across wide frequency ranges. 
Information on the predominant direction of propagation of candidate AF periods with respect to the local magnetic field direction is derived from the normalized cross helicity:
\begin{equation}
    \sigma_c(k,t)=\frac{W_{+}(k,t)–W_{–}(k,t)}{W_{+}(k,t)+W_{–}(k,t)},
\end{equation}
where $W_\pm(k,t)$ are the sum of the power of the wavelet transforms of the components of the Elsässer variables $\vec{z}_\pm(t) = \vec{v}(t) \pm \vec{v}_A(t)$, with $\vec{v}_A(t)$ being the Alfv\'en velocity of the plasma. 
The Elsässer variables represent a useful formalism to identify the dominant direction of propagation of AFs along a background magnetic field. $\sigma_c(k,t)$ is expected to be $<0$ ($>0$) for dominant propagation parallel (anti-parallel) to the local magnetic field direction, and $\sim 0$ for a balanced propagation along both directions. 

For each ICME, we perform the wavelet analysis using the Paul wavelet \citep[due to its better time localization capability than the Morlet wavelet;][]{Telloni2012} and considering a period of $2$ days before the ICME start and $2$ days after the ME end in order to avoid effects related to the cone of influence (the region of the wavelet spectrum where edge effects become important) at the edges of the time period of interest. 
We perform the analysis on both ACE and Wind data, and for Wind we consider two different plasma data sets, i.e. from the SWE and 3DP instruments, to check for instrumental/processing and temporal resolution effects that might affect the identification of AFs. 
Before applying the wavelet transforms to magnetic field and plasma time series, we resample them to a common cadence at both ACE and Wind. The resampling cadence is dictated by the largest cadence available across all data sets at both Wind and ACE, which in our case is 92-s based on Wind/SWE plasma data.

Based on these results, at each time step we integrate $\sigma_r(k,t)$ and $|\sigma_c(k,t)|$ across different scales by computing their median values across scales $k_i$ corresponding to periods between 5 min and 12 hours \citep[corresponding to $2.3 \times 10^{-5} - 3.3 \times 10^{-3}$ Hz, and falling within the injection range and the low-end of the inertial range of the power spectrum; see][]{Good2020}. The sample points used to determine these averages are equally spaced across the linear frequency range. In this way we obtain time-dependent medians for $\sigma_r(k,t)$ and $|\sigma_c(k,t)|$ which are purely functions of time $t$. 
Additionally, we also define an ``Alfv\'{e}nicity parameter'' as 
\begin{equation}
    p_A(t) = \mathrm{median}_k(|\sigma_c(k,t)|) (1-\mathrm{median}_k(|\sigma_r(k,t)|)),
\end{equation} 
which runs between 0 and 1 and measures the Alfv\'{e}nicity of the structure at time $t$. The closer $p_A(t)$ is to 1, the stronger the Alfv\'{e}nic content of the structure at time $t$. 
The quantification of Alfv\'{e}nicity at time $t$ through the Alfv\'{e}nicity parameter $p_A(t)$ assumes that AFs have a predominant direction of propagation, so that $\mathrm{median}_k(|\sigma_c(k,t)|)$ is close to 1. 

These three quantities ($\sigma_r$, $\sigma_c$, $p_A$) provide complementary metrics to evaluate the Alfv\'enicity of fluctuations within plasma.
While $\sigma_r$ enables the characterization of Alfv\'enicity through the identification of all AFs (both uni-directional and counter-propagating), the consideration of $p_A(t)$ specifically targets Alfv\'enic periods with predominantly uni-directional AFs. Finally, $\sigma_c$ identifies uni-directional AFs while not considering that AFs may actually provide only a negligible contribution compared to other wave modes. Considering $\sigma_r$ and $p_A$ (i.e.\ a combination of $\sigma_r$ and $\sigma_c$) therefore guarantees a more accurate representation of the actual contribution of uni-directional vs counter-propagating AFs within MEs.

\subsubsection{Correlation of multi-point magnetic field measurements} 
\label{subsubsec:methods_correlation}

In order to investigate the relation between a coherent behavior and the correlation of ICME signatures measured at different locations, we compute the correlation between the time profiles of the magnetic field strength and magnetic field components within the MEs measured at Wind and ACE. 
To do so, for each event we take the shock time and ME boundaries at Wind as references. 
The sheath and ME time series portions at ACE are each shifted and stretched to match the sheath and ME start and end times at Wind. 
We then analyze the correlation in two ways: 
(1) first, we compute the global correlation of the magnetic field strength and magnetic field components within the ICME between Wind and ACE.
In this case, both Wind and ACE data sets are rebinned to 30-min averages following the same approach of \citet{Lugaz2018}. The correlation is then computed separately for sheath and ME periods, for the magnetic field strength and the three magnetic field components, and is provided in the form of global Pearson correlation coefficients $\vec{cc} = (cc_B, cc_{B_R}, cc_{B_T}, cc_{B_N})$. This approach enables us to measure the global synchronicity between measurements taken at Wind and ACE, and reduces the relation between the two signals to a single value.\
(2) To gain insight into how the magnetic field correlation is distributed throughout the different ICME sub-structures, we further explore the instantaneous (i.e. time-dependent) Pearson correlation between ACE and Wind time series as a function of different time scales. 
We do so in both the sheath and ME regions by measuring the Pearson correlation between the Wind and ACE starting from a small portion of the signal, and then repeating the process along a rolling window until the entire structure is covered. 
To be consistent with the time scales explored in the study of Alfv\'enicity in Section~\ref{subsubsec:methods_af_identification}, we consider time windows $\Delta t_i$ equally spaced between 5 minutes and 12 hours with increments of 5 minutes, and integrate the results across these time scales by computing their median values ($\mathrm{median}_{\Delta t}$) across the various $\Delta t_i$ considered. 
The correlation of the magnetic field strength and three magnetic field components across the ME is provided in the form of time-dependent Pearson correlation coefficients $\vec{cc}(t) = (\mathrm{median}_{\Delta t}(cc_B(t)), \mathrm{median}_{\Delta t}(cc_{B_R}(t)), \mathrm{median}_{\Delta t}(cc_{B_T}(t)), \mathrm{median}_{\Delta t}(cc_{B_N}(t)))$.

\subsubsection{Superposed epoch analysis of Alfv\'enicity and magnetic field correlation} 
\label{subsubsec:methods_sea}

To determine the general profiles of Alfv\'enicity and time-dependent magnetic field correlation observed within ICME sheaths and MEs at 1 au, we make use of the superposed epoch analysis \citep[SEA;][]{Chree1913} technique. This technique allows the superposition of the time profile for a given quantity observed for different events, and the calculation of its averaged time profile. 
In the case of structures with multiple well-defined boundaries such as ICMEs (i.e., shock time, ME start, ME end), the time series are normalized in time for each sub-structure. Such three-bound SEAs have previously been performed to determine the average magnetic field and plasma profiles of ICMEs \citep[e.g.,][]{Masias-Meza2016, Regnault2020, Janvier2021} and the average Alfv\'enicity profile at inertial scales in terms of $\sigma_r(t)$ and $\sigma_c(t)$ \citep[e.g.,][]{Good2022}, but to the best of our knowledge, they have never been applied to multi-point correlation profiles within ICMEs.

In this work, we investigate the sheath and ME profiles of $\sigma_r(t)$, $|\sigma_c(t)|$, $p_A(t)$, and of $\vec{cc}(t)$ between ACE and Wind using a three-bound SEA with the sheath start, ME start, and ME end as reference times. The normalized time unit is set to be between 0 and 1 for the sheath region. Then, an average scaling factor is calculated as the ratio of the average ME duration to the average sheath duration, across all events considered. This average scaling factor is 2.18, which is used to determine the duration of the normalized time for the SEA ME profile. As a result, the normalized time runs from 0 to 1 for the sheath, and from 1 to 3.18 for MEs. From the normalized time series, the $\sigma_r(t)$, $|\sigma_c(t)|$, $p_A(t)$, and $\vec{cc}(t)$ data for each event are averaged into 22 bins within the sheath region, and 50 bins within the ME region, corresponding to bins of about 30 minutes in both the sheath and ME regions. 
We set bins of 30 minutes to ensure a sufficient resolution to resolve the temporal variation of the various SEA quantities across both sheaths and MEs, similarly to \citet{Good2022}.  
The mean and median values for each bin across all events are then calculated in order to build average profiles of $\sigma_r(t)$, $|\sigma_c(t)|$, $p_A(t)$, and $\vec{cc}(t)$ within ICMEs at 1 au when observed at longitudinal separations of $0.5^\circ-0.7^\circ$.

\section{Results}
\label{sec:results}

\subsection{Average Alfv\'{e}nic Content}
\label{subsec:statistics_alfvenicity}

\begin{sidewaysfigure}[ht!]
 \centering
 \includegraphics[width=\linewidth]{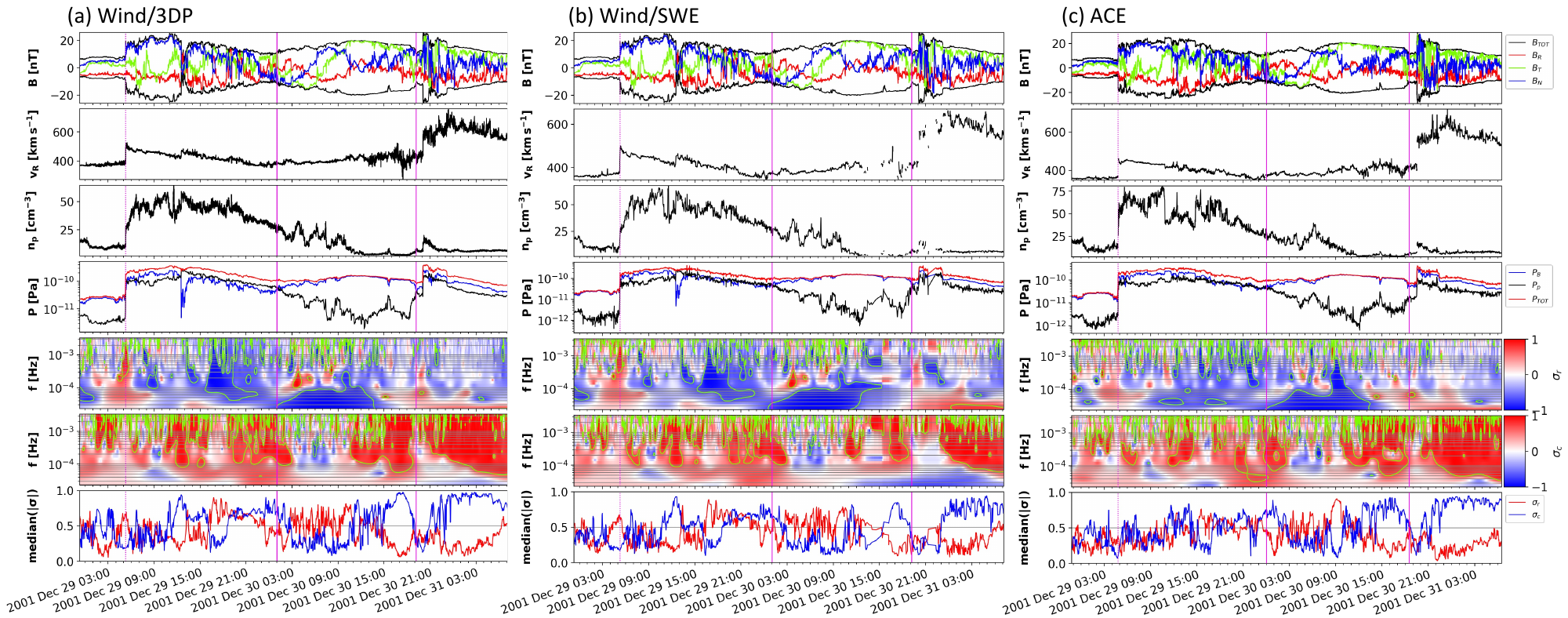}
\caption{
Comparison between Wind and ACE observations for an example event (event no. 7 in Table~\ref{tab:icmes}).
(a): Wind/MFI magnetic field and plasma data from Wind/3DP. 
(b): Wind/MFI magnetic field and plasma data from Wind/SWE.
(c): ACE/MAG magnetic field and plasma data from ACE/SWEPAM.
From top to bottom: magnetic field components; radial velocity; proton density; magnetic, thermal, and total pressures; 
$\sigma_r$; $\sigma_c$; median of $|\sigma_r|$ and $|\sigma_c|$ across scales $k_i$ between 5 min and 12 hours.
The vertical magenta lines indicated the ICME shock time and the ME start and end times.}
\label{fig:event1_insitu_comparison}
\end{sidewaysfigure}

As a starting point, we aim to characterize how many ICMEs, and what fraction of their sheath and ME exhibit a high Alfv\'enicity at Wind and ACE.

\smallskip
We start by evaluating the average Alfv\'enicity of ICMEs by computing the mean $\sigma_r$, $|\sigma_c|$, and $p_A$ and their standard error (SE) across the different events for the Wind/3DP, Wind/SWE, and ACE data sets. Results are provided in Table~\ref{tab:statistics_alfvenicity}.
We observe that the results from the different data sets are consistent within their estimated uncertainties, so overall, we find a good agreement across different instruments and spacecraft near 1 au. Differences among the different data sets can be traced back to instrumental and processing effects entering the measurement and extraction of ion moments (primarily the proton velocity and density in this case) which enter the calculation of the plasma Alfv\'enicity \citep[e.g.][]{King2005}. The comparison between Wind/3DP, Wind/SWE and ACE data for an example event is provided in Figure~\ref{fig:event1_insitu_comparison}. Despite having resampled all data sets to a common resolution, and despite our consideration of time scales larger than 5 min in this study, we point out that the different (intrinsic) time resolution between Wind/3DP and Wind/SWE measurements as well as data gaps that may be present in one of the two data sets may additionally contribute to slight differences in the resulting Alfv\'enicity calculation. 

\citet{Good2022} performed a similar analysis of sheaths and MEs for the frequency range $10^{-3}-10^{-2}$~Hz \citep[equivalent to wave periods of 16.7 -- 1.67 minutes, falling in the inertial range of the power spectrum;][]{Good2020}, reporting a mean $\sigma_r$ of $-0.36$ for MEs ($-0.35$ across sheaths) and a mean $\sigma_c$ (corrected for the sector magnetic polarity, so approximately corresponding to $|\sigma_c|$) of 0.18 for MEs (0.24 for sheaths). In the solar wind, \citet{Chen2013} found a mean $\sigma_r=-0.19$ and a mean $\sigma_c=0.40$.
The negative values found for $\sigma_r$ in this work is therefore consistent with previous works in finding that ME and sheath fluctuations exhibit a higher deviation from energy equipartition (in favor of magnetic fluctuations) compared to the ambient solar wind. The higher $|\sigma_c|$ retrieved in both sheaths and MEs compared to \citet{Good2022} also shows that Alfv\'enic fluctuations within ICMEs tend to be less balanced in their propagation direction at injection scales compared to inertial scales investigated by \citet{Good2022}.  
We chose to consider $|\sigma_c|$ in order to compare ICME events that have positive and negative $\sigma_c$ values. However, $|\sigma_c|$ only allows to distinguish between AFs that are uni-directional (high $|\sigma_c|$) vs counter-propagating (low $|\sigma_c|$), but does not allow us to draw conclusions on the specific direction of propagation of AFs with respect to the magnetic field background (parallel for $\sigma_c < 0$ and anti-parallel for $\sigma_c > 0$).
The usefulness of $\sigma_c$ to infer information on the AF origin and propagation for individual ICMEs will be demonstrated in an upcoming study currently in preparation.

\begin{table}[]
\centering
\begin{tabular}{ll|ccc}
 \hline
 \hline
  &                                             & Wind/3DP             & Wind/SWE           & ACE \\ 
 \hline
 Sheath: & $\sigma_r \pm \mathrm{SE}_{\sigma_r}$    & $ -0.35 \pm 0.04 $   & $ -0.22 \pm 0.04 $ & $ -0.30 \pm 0.03 $ \\ 
  & $|\sigma_c| \pm \mathrm{SE}_{|\sigma_c|}$       & $ 0.50 \pm 0.03 $    & $ 0.35 \pm 0.04 $  & $ 0.44 \pm 0.03 $ \\ 
  & $p_A \pm \mathrm{SE}_{p_A}$                     & $ 0.31  \pm  0.02$   & $ 0.23 \pm 0.02 $  & $0.29 \pm 0.02 $ \\ 
 \hline
 ME: & $\sigma_r \pm \mathrm{SE}_{\sigma_r}$        & $ -0.32 \pm 0.04$ & $-0.23 \pm 0.03$ & $-0.25 \pm 0.03$ \\ 
  & $|\sigma_c| \pm \mathrm{SE}_{|\sigma_c|}$       & $ 0.45 \pm 0.03$  & $0.34 \pm 0.03$  & $0.40 \pm 0.02$ \\ 
  & $p_A \pm \mathrm{SE}_{p_A}$                     & $ 0.30 \pm 0.03$  & $0.24 \pm 0.03$  & $0.27 \pm 0.02$ \\ 
 \hline
 \end{tabular}
 \caption{Mean $\sigma_r$, $|\sigma_c|$, and $p_A$ across the sheaths and MEs in our sample, for different data sets.}
\label{tab:statistics_alfvenicity}
\end{table} 

\begin{figure}[ht!]
 \centering
 \includegraphics[width=\linewidth]{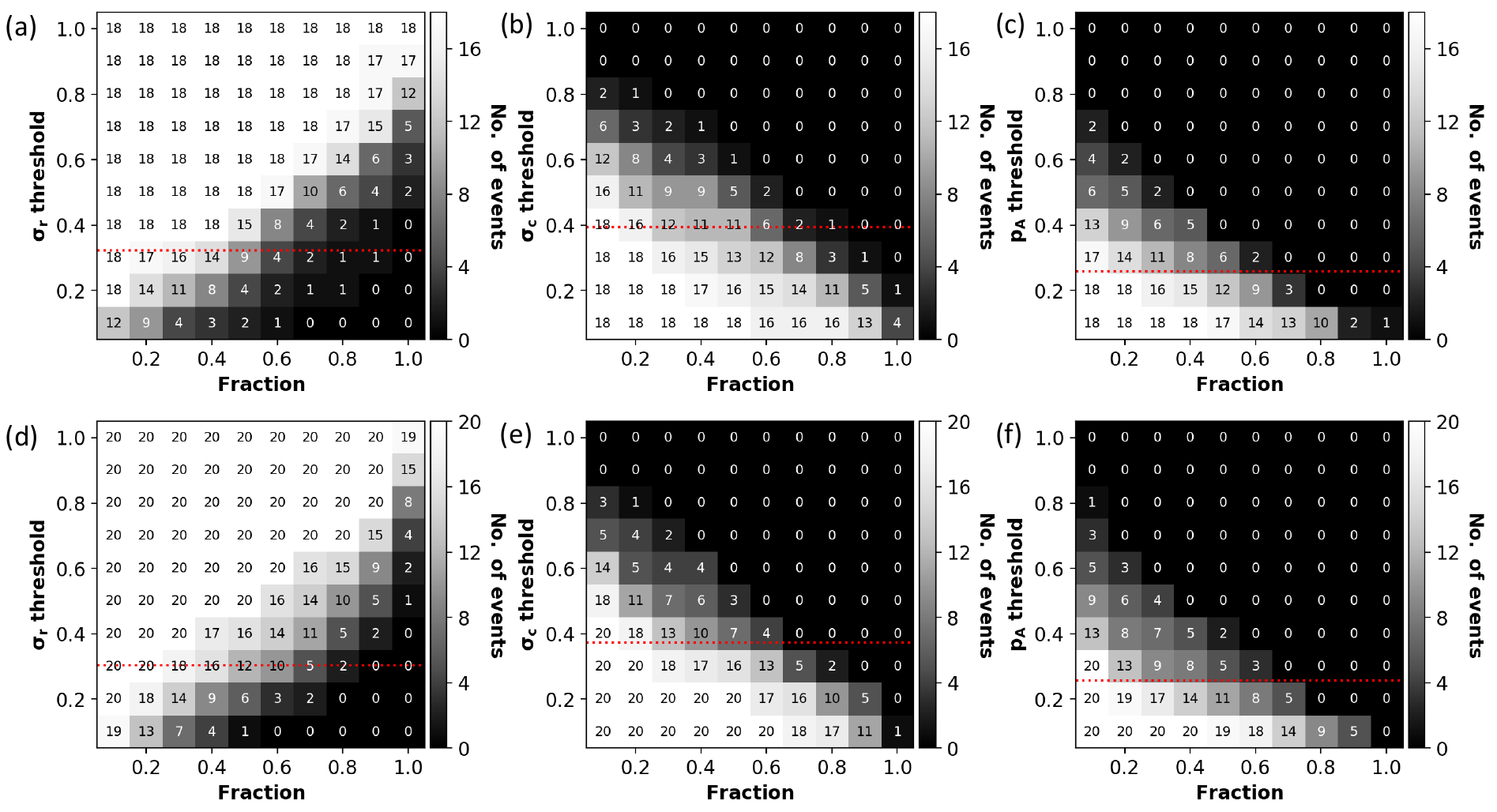}
\caption{
Alfv\'enic content of sheaths and MEs for different threshold values $\sigma_r^{*}$, $\sigma_c^{*}$, and $p_A^{*}$.
(a), (b), (c): combined results for sheaths from ACE and Wind/SWE.
(d), (e), (f): combined results for MEs from ACE and Wind/SWE.
The red dashed lines show the mean values reported in Table~\ref{tab:statistics_alfvenicity}.
}
\label{fig:statistics_alfvenicity}
\end{figure}

\smallskip
Next, we aim to establish to what extent ICMEs are Alfv\'enic.\
Answering this question requires a formal definition for the identification of AFs, particularly in terms of threshold values for $\sigma_r$, $\sigma_c$, and $p_A$, in order to discriminate between (highly) Alfv\'enic and non-Alfv\'enic periods. 
Since AFs show $\sigma_r \sim 0$ and uni-directional AFs are marked by $|\sigma_c| \sim 1$, strong AFs exhibiting a predominant direction of propagation are expected to be associated with $p_A \sim 1$. 
Because the thresholds to distinguish between Alfv\'enic and non-Alfv\'enic plasma are unknown, we explore different thresholds for $\sigma_r$, $\sigma_c$, and $p_A$, and evaluate how many MEs meet each of their respective threshold conditions. 
Specifically, we consider different threshold values $\sigma_r^{*}$, $\sigma_c^{*}$, and $p_A^{*}$ and evaluate how many events fulfill $|\sigma_r|$ lower than (or $|\sigma_c|, \,\, p_A$ greater than) each of these thresholds.  
We compute the number of MEs satisfying these conditions for different fractions of the ME duration, i.e.\ from 10\% to 100\%. Our results are shown in Figure~\ref{fig:statistics_alfvenicity} for both sheaths (panels (a) to (c)) and MEs (panels (d) to (f)) for a combination of Wind/SWE and ACE observations. As highlighted above, the results from Wind/3DP (not shown) are slightly different but they do agree with those from Wind/SWE and ACE within the uncertainties.

From Figure~\ref{fig:statistics_alfvenicity}~(d), we observe that the 
minimum value threshold satisfied by all MEs is $|\sigma_r| \le 0.2$, but this happens in all events for only 10\% of the ME duration. 
70\% of the ME observations (8 at Wind, and 6 at ACE) satisfy the condition $|\sigma_r| \le 0.2$ for at least 30\% of the ME duration. 
90\% of them (9 at Wind, and 9 at ACE) exhibit $|\sigma_r| \le 0.2$ for at least 20\% of the ME duration. 
Imposing a stricter Alfv\'enicity condition, we find that 65\% of the ME observations (8 at Wind and 4 at ACE) present levels of $|\sigma_r| \le 0.1$ for at least 20\% of the ME duration.
These results prove that the majority of ICMEs considered exhibit highly Alfv\'{e}nic conditions for a significant portion of their MEs, indicating AFs within MEs at 1 au may be more common than initially estimated \citep[e.g. by][]{Marsch2009, Yao2010}.
Sheaths are less Alfv\'enic than MEs (Figure~\ref{fig:statistics_alfvenicity} (a)), as indicated by the slightly higher mean $|\sigma_r|$ and by the lower number of events (6 for Wind, and 3 for ACE) exhibiting AFs ($|\sigma_r| \le 0.1$) for more than 20\% of the total duration. 

Results for the normalized cross helicity (Figure~\ref{fig:statistics_alfvenicity}~(e)) show that the
maximum value threshold satisfied by all MEs is $|\sigma_c| \ge 0.4$, but this is sustained in all events for only 10\% of the ME duration. 
Periods of $|\sigma_c| \ge 0.5$ are observed in 90\% of the cases, with a variable duration between 10\% and 50\% of the ME.
For comparison with the $|\sigma_r|$ signatures discussed above, we focus our attention to the maximum $|\sigma_c|$ found in most MEs for at least 20\% of their duration: this is measured to be $|\sigma_c| \ge 0.5$, and it applies to 55\% of the MEs.
The extent of the white areas in Figure~\ref{fig:statistics_alfvenicity} also provide evidence that while AFs are rather common in MEs (panel (d)), they often occur without a predominant direction of propagation (as indicated by the smaller white area in panels (e) and (f)). 
Similar conclusions hold for sheaths. However, they typically feature a higher $|\sigma_c|$ than MEs, both from an average standpoint and in terms of duration, indicating that slightly more unidirectional AFs are present in sheaths.

We combine signatures of low residual energy and high cross helicity together into the Alfv\'{e}nicity parameter in Figure~\ref{fig:statistics_alfvenicity}~(f). We observe that a smaller (though considerable) number of MEs satisfy the threshold condition $p_A \ge p_A^{*}$ than for the residual energy and cross helicity taken independently, as indicated by the smaller white area in panel (f) compared to panels (d) and (e).
We can further quantify the contribution from uni-directional vs.\ counter-propagating AFs by considering that for $\sigma_r^* = 0.1$ and $\sigma_c^* = 0.5$ (corresponding to the limit thresholds holding for at least 20\% of the ME duration in most of the events), we expect $p_A^* = \sigma_c^*(1-\sigma_r^*) = 0.45$. We obtain that 30\% to 40\% of MEs exhibit $p_A \ge 0.45$ for at least 20\% of their duration, compared to 65\% for $|\sigma_r| \le 0.1$, and 55\% for $|\sigma_c| \ge 0.5$. Based on these fractions, we estimate that between two thirds and one half of all AFs are uni-directional, while one third to one half are likely counter-propagating. The lower detection retrieved from the consideration of $p_A$ over $\sigma_c$ also implies that a significant fraction of all Alfv\'{e}nic periods identified through the cross helicity calculation may not actually be Alfv\'enic.
Sheaths tend to present slightly higher values of $p_A$ than MEs, but just as for MEs, this parameter shows that only a small fraction of AFs within sheaths propagate in a predominant direction within truly Alfv\'{e}nic periods.
We conclude that while the cross helicity may be used to reliably define Alfv\'enicity in the solar wind \citep[e.g.][]{Stansby2019}, it falls short in accurately representing Alfv\'enicity in ICME sheaths and MEs. In contrast, the Alfv\'{e}nicity parameter enables a more accurate representation of the actual contribution of uni-directional AFs within sheaths and MEs.

To the best of our knowledge, the only statistical survey making a quantitative estimation of the frequency and duration of AFs within MEs was performed by \citet{Li2016b}. The authors investigated 33 MEs observed by Voyager 2 between 1 and 6~au, and in contrast to the previous literature generally identifying a scarcity of AFs within MEs, \citet{Li2016b} also reported abundant AFs within MEs as we do in this study. Specifically, the authors considered scales from $5 \times 10^{-4}$~Hz to $10^{-2}$~Hz (corresponding to 33 to 1.7 minutes) and reported a highly Alfv\'enic content in about 90\% of the MEs investigated, and found AFs were present for about 17\% of the ME duration at 1.5 au. The fraction was also found to decrease with heliocentric distance, and could be back-extrapolated to about 20\% at 1 au assuming a linear decay with heliocentric distance. Their estimate at 1~au, although retrieved from a different methodology \citep[based on the Wal\'en test; see][]{Li2016}, is very consistent with the estimates reported from our study for a $|\sigma_r| \le 0.2$ threshold, and makes us confident of the robustness of our results. \citet{Li2016b} also found the duration of AFs within MEs to reduce to 4\% near 6~au. While the results by \citet{Li2016b} indirectly support a solar origin and interplanetary dissipation of AFs within ICMEs during propagation through larger heliocentric distances, the actual origin of AFs within ICMEs remains debated. A definitive conclusion remains difficult to achieve without contextualization of single-point observations with respect to the solar wind conditions encountered by individual ICMEs during propagation, and of the solar environment affecting their early evolution. 

\subsection{Superposed Epoch Analyses of Alfv\'{e}nicity and Magnetic Field Correlation}
\label{subsec:statistics_sea}

\begin{figure}[ht!]
 \centering
 \includegraphics[width=\linewidth]{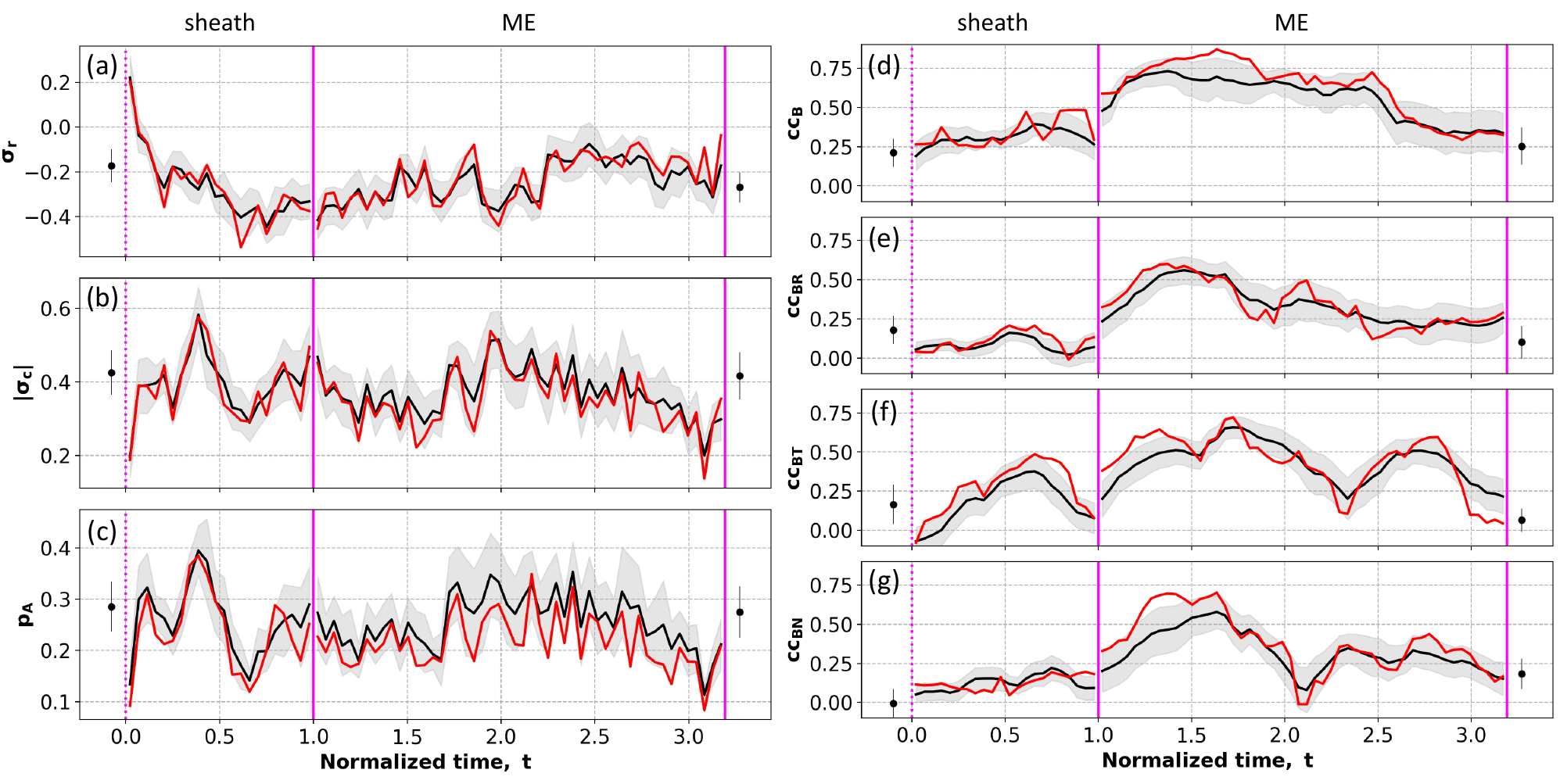}
\caption{Results of the SEA analysis. 
Panels (a), (b), and (c) show the profiles for the median $\sigma_r$, $|\sigma_c|$, and $p_A$ profiles for sheaths and MEs at ACE and Wind/SWE.
Panels (d), (e), (f), and (g) show the profiles for the correlation of $B$, $B_R$, $B_T$, and $B_N$ between ACE and Wind for sheaths and MEs.
The mean and median profiles are given by the black and red lines, respectively, with associated SEs reported as shaded areas. 
Single bins spanning 6 hours of solar wind before and after the ICMEs are also shown. Vertical dashed lines indicate the shock, start and end of the ME profile. 
}
\label{fig:statistics_sea}
\end{figure}

Next, we are interested in exploring the temporal distribution of Alfv\'{e}nicity and magnetic field correlation within sheaths and MEs at 1 au. To do so, we perform SEAs of the median $\sigma_r(t)$, $|\sigma_c(t)|$, and $p_A(t)$ time profiles at scales $k_i$ corresponding to time scales between 5 minutes and 12 hours. The combined results from ACE and Wind/3DP observations are shown in Figure~\ref{fig:statistics_sea} (panels (a) to (c)). 
As from the inspection of global Alfv\'{e}nicity metrics discussed above, here we find slightly negative $\sigma_r(t)$ values across both sheaths and MEs, which are consistent with \citet{Good2022} in indicating an excess of magnetic energy in the fluctuations within ICMEs. 
The temporal profile of $\sigma_r(t)$ also indicates that the Alfv\'enicity is higher immediately downstream of the shock and rapidly decreases throughout the sheath, suggesting Alfv\'enic fluctuations formed in the shock downstream region may efficiently decay into compressive modes or undergo damping phenomena as they propagate through highly turbulent sheaths \citep[see, e.g.,][]{Ala-Lahti2019, Farrugia2020}. Conversely, $|\sigma_c(t)|$ and $p_A(t)$ are larger in the first half of the sheath, suggesting that when present, AFs near sheath fronts tend to propagate along preferential directions. 
Throughout the ME, $\sigma_r(t)$ shows an increase towards zero:
7 out of 10 MEs exhibit a more negative average $\sigma_r$ in the ME first half than in the second half (on average more negative by $-0.09$ at Wind and by $-0.12$ at ACE). This suggests that AFs are preferentially located within ME backs than ME fronts. 
$|\sigma_c(t)|$ and $p_A(t)$ present irregular behaviors without a clear increasing or decreasing trend within MEs.
Their values exhibit minimal differences between the ME first and second half: on average, 
$|\sigma_c(t)|$ is only 0.01 larger in the ME first half, while $p_A(t)$ is equal in the ME first and second half at both ACE and Wind. This implies that the directionality of AFs within MEs is not related to their location within ME structures.
We also note that the SEA profiles of all the above parameters present a high temporal variability (both within sheaths and MEs), underlying a high temporal variability within individual event profiles.

\begin{table}[]
\centering
\begin{tabular}{ll|ccc}
 \hline
 \hline
                                               &    &  Global (30-min resample) & Global (92-s resample) & Time-dependent \\ 
 \hline
 Sheath: & $cc_B \pm \mathrm{SE}_{cc_B}$       &  $ 0.56 \pm 0.09$  &  $ 0.44 \pm 0.08$   & $ 0.31 \pm 0.06 $ \\ 
  & $cc_{BR} \pm \mathrm{SE}_{cc_{BR}}$        &  $ 0.33 \pm 0.11$  &  $ 0.19 \pm 0.05$   & $ 0.08 \pm 0.05 $ \\ 
  & $cc_{BT} \pm \mathrm{SE}_{cc_{BT}}$        &  $ 0.64 \pm 0.06$  &  $ 0.38 \pm 0.11$   & $ 0.18 \pm 0.09 $ \\ 
  & $cc_{BN} \pm \mathrm{SE}_{cc_{BN}}$        &  $ 0.50 \pm 0.08$  &  $ 0.24 \pm 0.07$   & $ 0.13 \pm 0.04 $ \\ 
 \hline
 ME: & $cc_B \pm \mathrm{SE}_{cc_B}$               &  $ 0.90 \pm 0.04$   & $ 0.88 \pm 0.05$    & $ 0.56 \pm 0.08 $  \\ 
  & $cc_{BR} \pm \mathrm{SE}_{cc_{BR}}$            &  $ 0.66 \pm 0.07$      & $ 0.59 \pm 0.08$    & $ 0.34 \pm 0.06 $ \\ 
  & $cc_{BT} \pm \mathrm{SE}_{cc_{BT}}$            &  $ 0.84 \pm 0.04$      & $ 0.79 \pm 0.05$    & $ 0.43 \pm 0.07 $ \\ 
  & $cc_{BN} \pm \mathrm{SE}_{cc_{BN}}$            &  $ 0.69 \pm 0.07$      & $ 0.64 \pm 0.07$    & $ 0.32 \pm 0.04 $ \\ 
  \hline
 \end{tabular}
 \caption{Mean magnetic field correlations between Wind and ACE for the sheaths and MEs in our sample, for time scales between 5 min and 12 hours.}
\label{tab:statistics_correlation}
\end{table} 

\begin{figure}[ht!]
 \centering
 \includegraphics[width=0.35\linewidth]{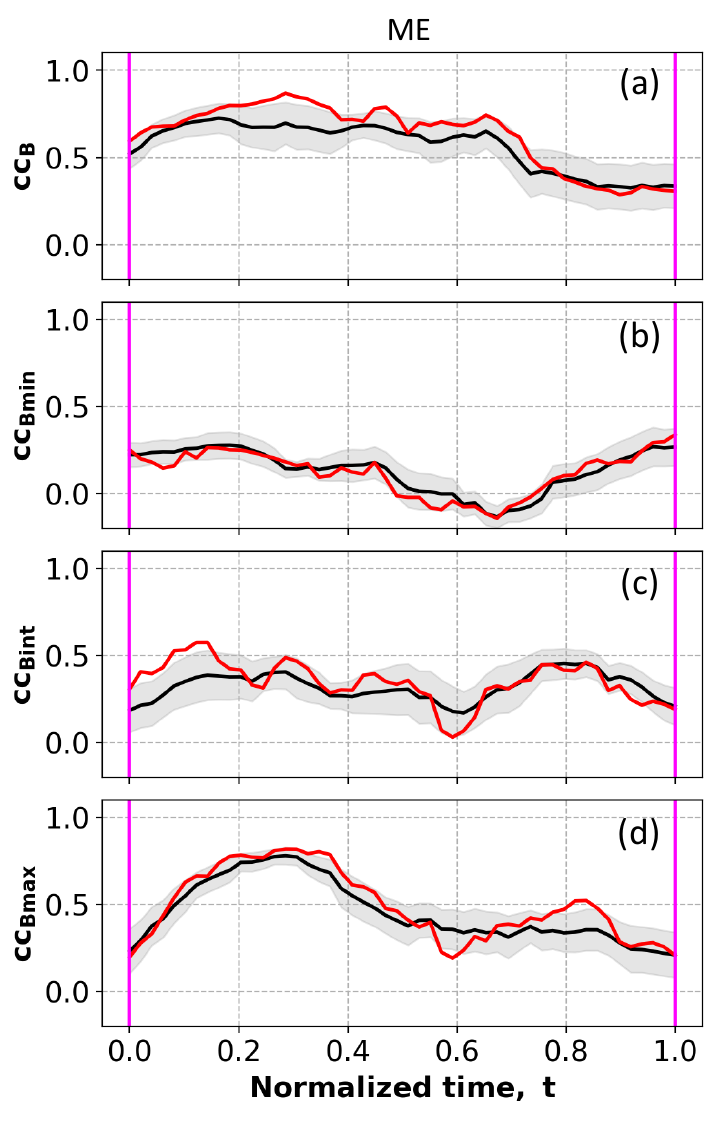}
\caption{SEA of the correlation of the ME magnetic field components, calculated after projecting each event into its MVA frame.
Panels (a), (b), (c), and (d) show the profiles for the correlation of $B$, $B_R$, $B_T$, and $B_N$ between ACE and Wind.
The mean and median profiles are given by the black and red lines, respectively, with associated SEs reported as shaded areas. 
Here the normalized time runs from 0 (start of the ME) to 1 (end of ME).
}
\label{fig:statistics_sea_mva}
\end{figure}

Second, we perform SEAs of the time profile of the correlation for each component of $\vec{cc}(t) = (cc_B(t), cc_{B_R}(t), cc_{B_T}(t), cc_{B_N}(t))$ between Wind and ACE for the ICMEs in our study. Figure~\ref{fig:statistics_sea} (panels (d) to (g)) shows the time-dependent distribution of correlation throughout the ICME sub-structures. 
We observe that the correlation in all magnetic field components tends to be higher near the ME front, while it is typically lower near the back. This is particularly evident for the total magnetic field, and for the $B_R$ and $B_N$ components. $B_T$ tends to have a more irregular profile without a clear decreasing or increasing trend throughout the ME. 
We can quantify this for individual events by comparing the average correlation in the first quarter and last quarter of each ME. We find that indeed, higher correlations are found at the ME front for 8 events in $B$ and $B_R$, 6 events in $B_N$, and only 3 events in $B_T$. On average, the ME front correlation is higher than the correlation in the ME back by 0.30, 0.21, 0.13, and 0.03 for $B$, $B_R$, $B_N$, and $B_T$, respectively. 
Because different MEs may have different orientations of their flux rope axis, we also apply the minimum variance analysis \citep[MVA;][]{Sonnerup1998} technique to each ME and repeat the analysis after having projected the magnetic field signatures of each event to that event's MVA frame. In this frame, approximately corresponding to the frame of the flux rope, the magnetic field components are projected in the $min$, $int$, and $max$ directions corresponding to the directions of minimum, intermediate, and maximum variance. For a flux rope structure, the $int$ direction corresponds to the direction of its magnetic axis, $max$ corresponds to the poloidal direction, and $min$ completes the right-handed triad. 
In terms of correlations (Figure~\ref{fig:statistics_sea_mva}), the MVA frames reveal that the largest correlation is present in $B_{max}$ (panel (d)), followed by $B_{int}$ (panel (c)) and $B_{min}$ (panel (b)). This trend is expected given that $B_{min}$ is likely to be the most sensitive component to the specific crossing of the spacecraft with respect to the flux rope axis, while $B_{max}$ is dominated by a large-scale bipolar signature that maximizes the correlation at different spacecraft over smaller-scale fluctuations \citep[see, e.g., Figure~2 in][]{DiBraccio2015}. $B_{int}$ is mainly uni-polar and presents an intermediate correlation between $B_{min}$ and $B_{max}$. 
Visually, in the MVA frame the only component exhibiting a clear difference between the ME front and back is $B_{max}$. As done in the RTN frame, we can therefore evaluate if this result holds for individual events by comparing the average correlation in the first quarter and last quarter of each ME. In this case we find that higher correlations are found at the ME front for 6 events in $B_{max}$ and $B_{min}$, and 5 events in $B_{int}$. On average, the ME-front correlation is higher than the ME-back one by 0.24, 0.10, $-$0.05 for $B_{max}$, $B_{min}$, and $B_{int}$, respectively. 

Sheaths appear less correlated than MEs but more correlated than the preceding and following solar wind in all magnetic field components, confirming the relative correlation scales identified by previous studies \citep[see][]{Wicks2010, Lugaz2018, Ala-Lahti2020}. 

For completeness, Table~\ref{tab:statistics_correlation} reports the average (mean) correlations obtained for sheaths and MEs observed at longitudinal separations of $0.5^\circ-0.7^\circ$ using the global \citep[as in][]{Lugaz2018} and the time-dependent correlation approaches described in Section~\ref{subsubsec:methods_correlation}. The results between the two methods are significantly different, with the time-dependent method consistently providing lower average correlation values than the global method. We argue that one reason behind this result is that the global method assumes a 30-min rebinning of the data, while the time-dependent one considers scales between 5 min and 12 hours. Repeating the global calculation using a rebinning of 92-s instead of 30-min, we obtain more similar results between the global and time-dependent calculations, especially within the sheath. 
In conclusion, while the time-dependent method represents a powerful and under-utilized tool to investigate the time evolution of ICME magnetic field correlations, its results can be significantly different from that of the more commonly-used global method. Therefore, in the perspective of future studies, it is critical to interpret the ICME magnetic field correlation coefficients with care and to only compare those obtained from similar methods.

\subsection{Correlation between Alfv\'{e}nicity and Magnetic Field Correlation}
\label{subsec:statistics_correlation}

\begin{table}[]
\centering
\begin{tabular}{ll|cc}
 \hline
 \hline
                        &           & Wind/SWE  &  ACE \\ 
 \hline
 Sheath: & $cc_B$ vs $\sigma_r$     & $ -0.84 $   (p$<$0.05)     & $ -0.79 $     (p$<$0.05)\\ 
  & $cc_{B_R}$  vs $\sigma_r$        & $ -0.14 $  (p=0.54)       & $ -0.19 $     (p=0.40)\\ 
  & $cc_{B_T}$  vs $\sigma_r$        & $ -0.73 $  (p$<$0.05)     & $ -0.69 $             (p$<$0.05)\\ 
  & $cc_{B_N}$  vs $\sigma_r$        & $ -0.66 $   (p$<$0.05)    & $ -0.65 $    (p$<$0.05)\\ 
 \hline
 ME: & $cc_B$ vs $\sigma_r$         & $ -0.35 $   (p$<$0.05)     & $ -0.22 $    (p=0.12)\\ 
  & $cc_{B_R}$  vs $\sigma_r$        & $ -0.42 $  (p$<$0.05)     & $ -0.21 $    (p=0.14)\\ 
  & $cc_{B_T}$  vs $\sigma_r$        & $ -0.12 $  (p=0.41)       & $ -0.13 $    (p=0.36)\\ 
  & $cc_{B_N}$  vs $\sigma_r$        & $ -0.13 $  (p=0.37)       & $ \,\,\,\, 0.10 $     (p=0.47)\\ 
  & $cc_{Bmin}$  vs $\sigma_r$        & $ -0.51 $ (p$<$0.05)     & $-0.51$      (p$<$0.05)\\ 
  & $cc_{Bint}$  vs $\sigma_r$        & $ \,\,\,\, 0.17 $  (p=0.23)      & $\,\,\,\, 0.06$      (p=0.67)\\ 
  & $cc_{Bmax}$  vs $\sigma_r$        & $ -0.33$  (p$<$0.05)    & $-0.12$      (p=0.41)\\ 
 \hline
 \end{tabular}
 \caption{Mean correlation between SEAs of $\sigma_r$ and the time-dependent magnetic field correlations at Wind and ACE for the sheaths and MEs in our sample, for time scales between 5 min and 12 hours.}
\label{tab:statistics_correlation_correlation_alfvenicity}
\end{table} 

After having characterized the  Alfv\'{e}nicity and magnetic field correlation independently of each other, we want to explore whether these two characteristics of ICMEs are correlated with each other and can provide insight into the role of AFs as mediators of coherent behavior across ICME structures. For this purpose, we calculate the correlation between $\vec{cc}(t)$ and $\sigma_r(t)$ obtained from the SEA profiles in Figure~\ref{fig:statistics_sea} within sheaths and MEs, and report the average values in Table~\ref{tab:statistics_correlation_correlation_alfvenicity}. 

We find that the correlation of the magnetic field profiles between the two spacecraft is weakly anti-correlated to the Alfv\'enicity of the MEs measured at each spacecraft (in terms of $\sigma_r$). 
The anti-correlation for $B$ is $-0.35$ and $-0.22$ for Wind/SWE and ACE, respectively.
For the magnetic field components, the anti-correlation ranges between $-0.12$ ($B_{T}$) and $-0.42$ ($B_R$) at Wind/SWE, and between $0.10$ ($B_N$) and $-0.21$ ($B_R$) at ACE.
However, as reported in Table~\ref{tab:statistics_correlation_correlation_alfvenicity}, some of these correlations are associated with p-values larger than 5\%, indicating we cannot reject the null hypothesis that the Alfv\'enicity and magnetic field correlation are not correlated. The most reliable (anti-)correlations are those for $B$ and $B_R$, which are associated with p-values smaller than 5\% at Wind  (while at ACE they present p-values around 10--15\%).
When comparing the Alfv\'enicity and the magnetic field correlations in the MVA frame, we observe a moderate anti-correlation in $B_{min}$ ($-0.51$ with p-value smaller than 5\% at both Wind/SWE and ACE), corresponding to the direction of minimum variance, and a weak anti-correlation for $B_{max}$ (from $-0.33$ with p$<5$\%, to $-0.12$ with p$=41$\% at Wind/SWE and ACE, respectively), corresponding to the direction of maximum variance. 
No statistically significant correlation/anti-correlation is found for $B_{int}$ ($0.17$ and $0.06$ with p-value larger than 5\% for Wind/SWE and ACE, respectively), corresponding to the direction of the flux rope axis. 
Because the magnetic field component aligned with the direction of propagation of AFs is expected to be the least affected by fluctuations, such a picture may suggest that AFs are globally propagating primarily along the flux rope axis \cite[along $B_{int}$, which typically constitutes the dominant magnetic field component within ICME flux ropes; see e.g.][]{Hu2015, Lanabere2020, Lanabere2022}, and manifest the larger anti-correlation in the other two magnetic field components ($B_{min}$ and $B_{max}$).

The anti-correlation is strong within sheaths (in terms of $\sigma_r$), where it ranges from $-0.84$ to $-0.14$ at ACE and $-0.79$ to $-0.19$ at Wind/SWE (results are similar for Wind/3DP). All correlations except the weakest ones associated with $B_R$ have an associated p-value that is smaller than 5\%, and can be thus considered statistically significant (Table~\ref{tab:statistics_correlation_correlation_alfvenicity}).
The stronger anti-correlation found within sheaths compared to MEs suggests that AFs may generate magnetic field differences at smaller scales within sheaths, while within MEs, the differences may affect larger scales. This arises from the consideration that fluctuations observed at different spacecraft may have different properties and be measured during different phases of their oscillation, and the smaller their scale and the larger the separation between the spacecraft measurements, the less correlation in the magnetic field signatures is observed at different spacecraft.
Another factor likely to increase the anti-correlation relates to the mechanism of formation of ICME sheaths \citep{Kilpua2017}. While the front part of sheaths is typically composed of shocked solar wind material, inner layers can be composed of preexisting compressed material of either coronal or interplanetary origin \citep{Vourlidas2013, Lugaz2020}, or even more coherent material originally part of the CME erupted structure which was later eroded, e.g., via magnetic reconnection processes at the ME front \citep[e.g.,][]{Dasso2006, Ruffenach2012, Ruffenach2015}.
At least in the case of ICME sheaths, the mixing of material of various origin (presenting different intrinsic Alfv\'enicity and correlation scales) likely contributes to the enhancement of the anti-correlation between the Alfv\'enicity and correlation of magnetic field components. Whether the strength of this anti-correlation can be used to infer information about the formation history of individual ICME sheaths remains unclear, but is certainly worth to be explored in future studies.

%
It is also noteworthy that these anti-correlations apply to the average sheath and ME profiles determined from the consideration of all 20 ICMEs profiles measured at ACE and Wind and analyzed through the SEA technique. Conversely, such a behavior is not found when taking the average of the correlations obtained from individual profiles contributing to the SEA (where we obtain correlations ranging between $0.00$ and $-0.18$ ($0.01$ to $-0.15$) for sheaths, and $-0.06$ to $-0.20$ ($0.13$ to $0.00$) for MEs at Wind/SWE (ACE); results are similar for Wind/3DP). In this respect, the SEA profiles reveal trends that are not immediately visible in individual events. The reason behind this is that individual events alternate periods of anti-correlation between $\sigma_r$ and the magnetic field correlation, with periods where their (anti-)correlation is not well determined. 
Additionally, the SEA profiles for $\sigma_r$ shown in Figure~\ref{fig:statistics_sea}~(a) shows a high temporal variability, indicating that individual events also present high temporal variability of this quantity, which likely makes the correlation with the (smoother) profiles in Figure~\ref{fig:statistics_sea}~(right) less evident when events are considered individually. 

As discussed in Section~\ref{subsec:statistics_alfvenicity}, highly Alfv\'{e}nic periods are found in most MEs but typically cover only 20\%--30\% of the ME duration. The anti-correlation between the Alfv\'{e}nicity and the correlation of the magnetic field from SEA profiles suggests AFs may contribute to reduce the correlation scale within sheaths and MEs, but that other factors may play a role as well. This appears particularly the case for MEs, which exhibit less clear anti-correlation trends than sheaths.

The slight preference of AFs for ME back regions (Figure~\ref{fig:statistics_sea} (a)), and the fact that lower magnetic field correlations are also found in that region (Figure~\ref{fig:statistics_sea} (right panels)) may also be explained in terms of AFs being formed, in most cases, through interaction of MEs with the following solar wind \citep[see for example the study by][]{Dhamane2023}. Conversely, the interaction with the preceding solar wind (even if highly Alfv\'enic) may not lead to an increase in the Alfv\'enicity within MEs due to the presence of the sheath, which may act to protect MEs from the propagation and formation of AFs \citep[this scenario is consistent with the case presented by][]{Farrugia2020}. 
Specifically with respect to the Alfv\'enicity time profile, every event was quite unique when considered alone. Nevertheless, we would like to mention that Event~6 (the only one in our set that did not drive a preceding shock and sheath) presented a relatively high Alfv\'enicity ($|\sigma_r| \sim 0.3-0.4$) at the ME front, and a lower Alfv\'enicity at the ME back ($|\sigma_r| > 0.5$) in ACE data. Such trend, however, was not clear from Wind observations, where the Alfv\'enicity tended to fluctuate around $|\sigma_r| \sim 0.5$ across the ME. Overall, it is hard to say if this is related to the lack of a preceding shock/sheath or to other factors (such as the eruptive scenario, or the propagation through interplanetary space), but taken on its own, this event seems to confirm our interpretation that the location of AFs of interplanetary formation may be heavily affected by the presence/absence of a preceding shock and sheath.

\section{Discussion and Conclusions}
\label{sec:conclusions}


In this work, we investigated the in-situ characteristics of 10 ICMEs (9 of which drove a shock and sheath) observed at 1 au by ACE and Wind while at longitudinal separations between $0.5^\circ$ and $0.7^\circ$ (corresponding to $\sim 200 - 300$ Earth radii, or $\sim 0.009 - 0.013$~au at 1~au).
For each event, we analyzed the Alfv\'{e}nicity of the sheath and ME in terms of the residual energy ($\sigma_r$) and cross helicity ($\sigma_c$) of fluctuations in the injection range and low-end of the inertial range at $2.3 \times 10^{-5}$ to $3.3 \times 10^{-3}$~Hz (corresponding to time scales between 5 min and 12 hours). We purposely considered such scales in order to explore the role of AFs in altering the internal structure of ICMEs at large to intermediate (i.e., ``meso'') scales. Additionally, we evaluated the coherence of ICMEs in terms of the correlation between the magnetic field signatures measured at Wind and ACE within sheaths and MEs.
The analysis of this set of 10 ICMEs highlighted the following trends:

\begin{itemize}

 \item The average Alfv\'enicity of ICME sheaths and MEs is comparable, and is broadly consistent when investigated using plasma data from the ACE, Wind/SWE, and Wind/3DP data sets (Figure~\ref{fig:event1_insitu_comparison} and Table~\ref{tab:statistics_alfvenicity}). 
 
 \item Though less common than in the solar wind, AFs are abundant within ICME sheaths and MEs (Figure~\ref{fig:statistics_alfvenicity}). Strongly Alfv\'enic periods ($\sigma_r \le 0.2$) lasting at least 20\% of a given sub-structure duration are found in about 65\% of sheaths and 90\% of MEs.

 \item Highly Alfv\'enic periods within sheaths and MEs exhibit a variety of $\sigma_c$ signatures, which provides information about the direction of propagation of AFs. About half of these highly Alfv\'enic periods are associated with AFs propagating either parallel or anti-parallel to the local magnetic field, while the other half of AFs are characterized by counter-propagating wave packets (Figure \ref{fig:statistics_alfvenicity}).
 
 \item From the construction of average time profiles of Alfv\'enicity within sheaths and MEs using the SEA method, we find that AFs within sheaths are mainly located immediately downstream of the shock, while AFs within MEs are preferentially located near the back of MEs (Figure~\ref{fig:statistics_sea} (a)).

\item When measured at longitudinal separations of $0.5^\circ-0.7^\circ$, the magnetic field profiles within sheaths are significantly less correlated than those within MEs (Figure~\ref{fig:statistics_sea}~(a) and Table~\ref{tab:statistics_correlation}).
The correlation is uniformly distributed within sheaths, while the correlation presents a decreasing trend throughout ME structures: ME backs tend to be less correlated than ME fronts, both in individual events and in the combined SEA profiles (Figure~\ref{fig:statistics_sea}~(b)).

 \item The comparison of the average profiles constructed through the SEA method reveals the Alfv\'enicity (measured in terms of $\sigma_r$) is anti-correlated to the magnetic field correlation. This anti-correlation appears stronger within sheaths, while it is weaker within MEs. Such an anti-correlation was not always prominent in individual ICMEs, and was more clearly revealed by the SEA.

\end{itemize}

Our first goal was to quantify the frequency and duration of AFs within ICMEs at 1 au.\
Having determined that AFs are abundant within sheaths and MEs, our second goal was to determine whether AFs can be responsible for mediating coherent behavior across ME structures, assuming such a behavior can be measured in terms of the correlation of magnetic field profiles measured at a given longitudinal separation (i.e.\ the higher the correlation, the higher the coherence). In other words, we tested if AFs can contribute to making ICMEs more self-similar along different directions (i.e. whether they increase the ICME magnetic field correlation scale). 
Contrary to early studies, but consistently with a statistical study by \citet{Li2016b} at larger heliocentric distances, we found that AFs are relatively abundant within ICME sheaths and MEs at 1 au. Surprisingly, our analysis suggests that instead of increasing the correlation of the magnetic field components within ICMEs, AFs may actually decrease it. 
This is in agreement with the fact that spacecraft crossing the same ME along different trajectories likely sample AFs in different oscillation phases, and can be interpreted in two ways. 
First, as an indication that the assumption that information mediating coherence across an ICME propagate at the Alfv\'{e}n speed (i.e. it is carried by Alfv\'{e}n waves) may be inappropriate. This in turn would imply that the correlation length may be even smaller than predicted by \citet{Owens2017} and \citet{Owens2020}, because information would necessarily propagate at slower speeds than the Alfv\'{e}n speed throughout MEs. In this case, a different carrier of information, alternative to Alfv\'{e}n waves, has to be identified in the future. 
Alternatively, in view of the anti-correlation between AFs and magnetic field correlation, one has to admit the possibility that the magnetic field correlation may not be a true measure of coherence, and that a re-thinking of the way we evaluate coherence based on in-situ data may be necessary. In this scenario, AFs would actually be mediators of information in the form of anti-coherence, and thus potentially in competition with mechanisms mediating coherence, (so far) still to be identified.
Large- to meso-scale waves would perturb the quasi-static (background) ICME structure simply due to their propagation. The larger their amplitude and the more separation between the spacecraft measurements, the less coherence will be observed. Changes to the ME structure may be either permanent or temporary due to the propagation of such waves through the structure.
The contextualization of single-point observations with respect to the solar wind conditions encountered by individual ICMEs during propagation, and of the solar environment affecting their early evolution is critical to further clarify these points.  We intend to demonstrate the close physical relationship between AFs and the correlation of magnetic field signatures of ICMEs in a following paper investigating a case study of one of the events from this larger study. 

Our analysis also provides indications about the possible evolution of the magnetic field correlation scales of MEs during propagation. 
The fact that despite interacting with the preceding and following solar wind, MEs still retain a larger correlation scale than the solar wind by the time they reach 1~au implies that MEs most likely have even larger correlation scales during the eruption/early propagation phases. One possible mechanism to explain such a degradation, as identified from our study, is given by AFs, which can have two distinct origins and effects on the magnetic field correlation profile of MEs. 
On the one hand, about one half of the identified AFs within MEs present counter-propagating in situ signatures (i.e. low $\sigma_r$ and low $p_A$) consistent with an origin at the Sun before the Alfv\'en surface. While the localization of these AFs within ME structures went beyond the capabilities of our analysis (as the $p_A$ SEA profile was designed specifically to locate uni-directional AFs rather than counter-propagating ones), such AFs are not expected to have a preferential location within ME cross sections, e.g., front, middle, back.
On the other hand, the results above indicate about one half of AFs within MEs propagate in a predominant direction (i.e. low $\sigma_r$ but high $p_A$), consistent with an interplanetary origin. These AFs are mainly found near ME backs.
This is a robust result holding for individual MEs, and suggests that two populations of AFs can co-exist within the general ME population at 1~au: one population originated at the Sun, i.e. prior to the CME crossing the Alfv\'en surface, which most likely had time to travel across the ME structure by the time it reaches 1~au \citep{Good2022} and to reduce the correlation scale uniformly across the whole ME cross section. And one population originated in the interplanetary space, primarily at the back of MEs through interaction with the following solar wind, which had time to propagate only locally within causally-connected ME regions and contributed to reduce the correlation length mainly near ME backs.
This interpretation is also consistent with the presence of a sheath (found in 90\% of the ICMEs in our set) that may ``protect" the front of the ME from being disrupted by the interaction with the preceding solar wind. Past studies suggested this may be the case particularly when the ICME propagates through a preceding highly Alfv\'enic solar wind \citep{Farrugia2020}. ME backs, on the other hand, would remain exposed to interactions with the following wind which may lead to the formation of AFs regardless of the presence/absence of a sheath ahead. This is especially true for MEs being overtaken by a following high speed stream. Fast solar wind streams are typically ``inundated'' by AFs \citep{Bruno2013} and can provide an additional reason why AFs in ICMEs tend to be more common at the back.

We conclude by emphasizing that these new insights have been obtained from the analysis of a small-population set of ICMEs near 1 au, due to past data limitations. Despite the relatively small sample size, through our analysis we were able to pinpoint the existence of significant trends that contribute to the understanding of the fundamental physical relationships between Alfv\'{e}n waves and ICME coherence. However, in the near future it will be of prime importance to validate these results through statistical studies considering larger sets of events, and through the examination of different heliocentric distances, in order to draw general conclusions regarding the relationship between Alfv\'enicity and the magnetic field correlation and coherence of ICMEs. 
The results presented in this study also serve as an important benchmark for further investigation and interpretation of individual ICME events, which we plan to address in an upcoming study currently in preparation. 
Finally, determining whether the correlation scales of the solar wind and ICMEs are larger closer to the Sun, what their relative magnitude is, and how quickly they drop with heliocentric distance will be important to understand how the solar wind can degrade the correlation scale of ICMEs in earlier propagation phases. Both statistical and case studies closer than 1 au are needed in order to characterize all these factors, and will soon be possible thanks to multi-point coordinated ICME observations from Parker Solar Probe and Solar Orbiter.

\begin{acknowledgments}
C.S. was supported by the NASA ECIP program (grant no. 80NSSC23K1057).
C.S. and N.L. acknowledge support from NASA grants 80NSSC20K0197 and 80NSSC20K0700.\ 
C.S. and R.M.W. acknowledge support from NASA grant 80NSSC19K0914.\
C.J.F.\ acknowledges support from NASA grant 80NSSC19K1293.\
N.M.\ acknowledges Research Foundation – Flanders (Fonds voor Wetenschappelijk Onderzoek (FWO) - Vlaanderen) for their support through Postdoctoral Fellowship 12T6521N.\
F.B.\ acknowledges support from the FED-tWIN programme (profile Prf-2020-004, project ``ENERGY'') issued by BELSPO.\
All data used in this study is publicly available through the NASA Solar Physics Data Facility (SPDF).\
The authors thank the ACE/SWEPAM, ACE/MAG, Wind/3DP, Wind/SWE, and Wind/MAG instrument teams for providing the necessary data to the public.
\end{acknowledgments}

%






\bibliography{camilla}{}
\bibliographystyle{aasjournal}



\end{document}